\documentclass[prl,twocolumn, reprint,nofootinbib, amssymb]{revtex4}
\usepackage{graphicx}
\usepackage{color}
\def\simlt{\lower.5ex\hbox{$\; \buildrel < \over \sim \;$}}
\def\simgt{\lower.5ex\hbox{$\; \buildrel > \over \sim \;$}}
\def\simpropto{\lower.2ex\hbox{$\; \buildrel \propto \over \sim \;$}}
\begin{document}

\title{Strangelets and the TeV-PeV cosmic-ray anisotropies}

\author{Kumiko Kotera$^{1,}$\footnote{kotera@iap.fr}, M. Angeles Perez-Garcia$^{2,}$\footnote{mperezga@usal.es},  and  Joseph Silk$^{1,}$\footnote{silk@iap.fr} }

\affiliation{$^1$Institut d'Astrophysique de Paris, UMR 7095 - CNRS, Universit\'e Pierre $\&$ Marie Curie, 98 bis boulevard Arago, 75014, Paris, France\\
$^2$Department of Fundamental Physics and IUFFyM, University of Salamanca, Plaza de la Merced s/n 37008 Salamanca}

\date{\today}

\begin{abstract}
Several experiments (e.g., Milagro and IceCube) have reported the presence in the sky of regions with significant excess in the arrival direction distributions of Galactic cosmic rays in the TeV to PeV energy range. Here we study the possibility that these hotspots are a manifestation of the peculiar nature of these cosmic rays, and of the presence of molecular clouds near the sources. We propose that stable quark matter lumps or so-called {\it strangelets} can be emitted in the course of the transition of a neutron star to a more compact astrophysical object. A fraction of these massive particles would lose their charge by spallation or electron capture in molecular clouds located in the immediate neighborhood of their source, and propagate rectilinearly without decaying further, hence inducing anisotropies of the order of the cloud size. With reasonable astrophysical assumptions regarding the neutron star transition rate, strangelet injection and neutralization rates, we can reproduce successfully the observed hotspot characteristics and their distribution in the sky.
\end{abstract}

\maketitle
Several experiments have reported strong anisotropy measurements in the arrival direction distributions of Galactic cosmic rays (CRs) in the TeV to PeV energy range (Super-Kamiokande, Tibet III, Milagro, ARGO-YBJ, and IceCube \cite{neutrino_exp,Abbasi11}). The data reveal the presence of large scale anisotropies of amplitude $\sim 0.1\%$. Smaller scale anisotropies of size $\sim10^\circ-30^\circ$ are also detected with amplitude a factor of a few lower. Milagro has reported the detection at significance $>12\sigma$ of two {\it hotspots} (regions with enhanced CR intensity) with amplitude $\approx 10^{-4}$, at a median energy of $1\,$TeV. ARGO-YBJ report similar excesses. IceCube observes localized regions of angular scale $\sim 15^\circ$ of excess and deficit in CR flux with significance $\sim5\sigma$ around a median energy of 20\,TeV~\cite{Abbasi11}. 

The large scale anisotropy could be naturally explained by the diffusive transport of CRs within the Galactic magnetic fields \cite{large_scale_theories,previous_works}. On the other hand, the intermediate and small scale anisotropies are more difficult to explain. The main difficulty resides in the fact that the Larmor radius of particles in the  TeV-PeV range is:
$r_{\rm L} \approx {E}/{ZeB} \sim 1.08\,{\rm pc}\,Z^{-1}(E/1\,{\rm PeV})(B/1\,\mu{\rm G})^{-1}$,
where the magnetic field strength of the Galaxy is assumed to be $B=1\,\mu$G (see \cite{Han08} for a review). For particles with $r_{\rm L}\ll l_{\rm c}$, where $l_{\rm c}=10-100\,\rm pc$ is the coherence length of the Galactic magnetic field (e.g., \cite{Han08}), the propagation will be totally diffusive over a  distance $>l_{\rm c}$. 
Neutrons would propagate rectilinearly, but their decay length around 10~TeV energies is less than 0.1\,pc. 
These scales are far shorter than the distance of any close-by source capable of accelerating particles to PeV energies. Various phenomena, such as heliospheric modulation, neutron sources, nearby pulsars, peculiar structures of the local Galactic magnetic fields have been invoked, but none seem to give an obvious explanation \cite{previous_works}. 

In this work we study the possibility that the hotspots in the skymap are a manifestation of the peculiar nature of CRs, and of the presence of molecular clouds (MCs) near the sources. We propose that quark matter lumps or so-called {\it strangelets} could be produced and accelerated while a neutron star (NS) transitions to a quark star (QS). A fraction of these heavy particles would suffer spallation or electron capture in molecular clouds located in the immediate neighborhood of their source, and produce neutral fragments that would propagate rectilinearly without decaying further, hence inducing anisotropies of the order of the cloud size. 

\section{Strangelet properties and sources}\label{properties}

Strangelets (also referred to as nuclearites) are supposed to be lumps of $uds$ quark matter. According to the Witten hypothesis \cite{witten84}, $ud$ matter is metastable and $d$-quarks decay by weak interaction, $u+d\rightarrow u+s$, to form more stable $uds$ matter. These lumps could be formed in explosive events like a NS undergoing a phase transition to a QS as proposed in dark matter (DM)-driven scenarios \cite{Perez-Garcia} or in the high-density environments of a compact object merger event \cite{merger}.
Direct searches are being conducted by, e.g., ground-experiments at the LHC (Alice and CMS experiment with the CASTOR calorimeters) or in space with the AMS-02 spectrometer. 

The mass number, $A$, of a stable strangelet can potentially range from a $\sim 10$  to $A\gg 10^{10}$. When considering a general non-zero strangeness-content, there is a nuclear stability valley for strangelets with baryonic number $A$ and there is a poorly known minimum value of mass number $A_{\rm min}\approx 10-600$ \cite{Wilk96} below which they are unbound. Typical values of strangelet binding energy are currently uncertain but supposed to be $E/A \sim \rm MeV-GeV$ energies. 
There is not much information on their possible charge value $Z$ but it should be small and (most likely) positive for finite lumps \cite{Madsen}. Several models of strangelets exist that lead to various $Z/A$ dependencies. For example, for ordinary strangelets, $Z=0.3A^{2/3}$, while for CFL (color-flavour-locked) strangelets $Z\simeq 0.3 A^{1/3}$ \cite{Madsen}. Even smaller charge-to-mass ratios are allowed $Z/A\sim -10^{-2}-10^{-7}$ . Experiments such as CREAM and AMS-02 will have the ability to perform a direct measurement of the charge, and infer estimates of $Z/A$. 

If strangelets were responsible for the  observed hotspots, they should produce detectable air-showers. This is possible if the kinetic energy per nucleon content, $K_{N}$, satisfies $K_N=K_{\rm tot}/A>1\,$GeV. Measurements indicate a total kinetic energy of particles in hotspots, $K_{\rm tot}\sim E \sim$\,TeV-PeV, which implies $A\lesssim 10^2-10^4$.

Neutron stars have been suggested as possible accelerators of strangelets \cite{strangelet_accelerators}. Strangelets could be produced for instance in the course of a NS to QS transition \cite{strange_stars}. In such events, a fraction $f_{\rm ej}$ of the gravitational energy released can be injected into the expelled outer crust, leading to total kinetic energies $E_{\rm ej}\sim 4\times10^{50}(f_{\rm ej}/10^{-3})\,$erg for standard NS mass and radius \cite{Perez-Garcia12}. The Lorentz factor of the ejected mass can be of order $\Gamma\sim 22\,(f_{\rm ej}/10^{-3})(12\,{\rm km}/R_{\rm *})(M_{\rm *}/1.5M_\odot)^2(10^{-5}M_\odot/M_{\rm ej})$, for NS mass $M_{\rm *}$, radius $R_{\rm *}$, and ejected mass $M_{\rm ej}$ \cite{Perez-Garcia12}. Particles of mass number $A$ could then gain energies of order $E_{\rm acc}\sim 21\,(A/10^3)(\Gamma/22)\,{\rm TeV}$, the typical energy observed in hotspots.

Accelerated stranglets may experience energy losses by interacting with the radiation field close to the NS, and with the baryonic and radiative backgrounds of the supernova (SN) envelope. Refs.~\cite{cosmic_rays_pulsars} concluded that there is room for the escape of accelerated particles. The discussion can be adapted to our case, except that strangelets are likely to have higher binding energies \cite{Madsen}, which will further help the escape. Besides, old NS may have a higher chance to undergo a transition \cite{Perez-Garcia12,perez-daigne-silk}, negligible radiative fields and no surrounding SN envelopes.

\section{Interaction with molecular clouds and Hotspots characteristics}

Once they have escaped from the source, strangelets diffuse in the magnetized interstellar medium (ISM). The trajectory of strangelets should be totally diffusive, even when assuming a low charge. They can reach the Earth on a timescale $\Delta t=d_{\rm s}^2/(2D)\sim 6\times 10^5\, Z^{1/3}(d_{\rm s}/1\,{\rm kpc})^2(E/{\rm 20\,TeV})^{-1/3}\,{\rm yrs}$, where $d_{\rm s}$ is the distance to the source and the diffusion coefficient is set to $D(E) = 1.33\times10^{28}H_{\rm kpc}[E/(3Z\,{\rm GeV})]^{1/3}$\,cm$^2\,$s$^{-1}$, with $H_{\rm kpc}\equiv H/(1\,{\rm kpc})$ the height of the Galactic halo \cite{Blasi12_1}. 
The ionization and the spallation timescales in the ISM (of average density $n_{\rm ISM}=0.5\,{\rm cm}^{-3}$) read respectively $\tau_{\rm ion}\sim 7\times10^{12}\,Z^{-2}(E/{\rm 20\,TeV})\,$yrs, and $\tau_{\rm spall}\sim 4\times 10^5\,(A/10^3)^{-2/3}\,(n_{\rm ISM}/ 0.5\, {\rm cm^{-3}})^{-1}\,$yrs \cite{Madsen}, implying that spallation should affect particles only mildly during their flight from sources located within 1\,kpc. 

Spallation could however play a prominent role if the source is born in or near a molecular cloud. Molecular clouds are the densest regions of  the ISM, and consist mainly of molecular hydrogen. Their typical radius in the Galaxy is $R_{\rm MC}\sim 20-50\,$pc, and their density $n_{\rm MC}\sim10^{2-6}\,{\rm cm}^{-3}$. In such regions, the spallation fraction can exceed unity, reaching  $r_{\rm spall}=\tau_{\rm esc}/\tau_{\rm spall}\sim 7.5\,Z^{1/3}(R_{\rm MC}/25\,{\rm pc})(n_{\rm MC}/10^3\,{\rm cm}^{-3})(A/10^3)^{2/3}$, with $\tau_{\rm esc}$ the diffusion time of strangelets in the cloud.   The electron capture rate for strangelets in clouds with free electron density $\sim \eta_e n_{\rm MC}$ (with $\eta_e\ll 1$) should be a fraction of the ionization rate, of order $r_{\rm ion}\sim 10^{-5}Z^{7/3}\eta_e (n_{\rm MC}/10^3\,{\rm cm}^{-3})$. As strangelets are predicted to be more bound than standard nuclei, these estimates can be viewed as upper limits for spallation. For electron capture, it is possible that the large size of strangelets dominates the effects of the charge, implying a scaling in $\sim A^{2/3}$, and the rates quoted here can be viewed as a lower limit. 

A fraction of strangelets undergoing spallation or electron capture (similar to that quoted for regular ions \cite{Padovani09}) may generate neutral secondaries. The work of \cite{Madsen} suggests that a tiny parameter space exists where spallation could lead to bound neutral strangelets.
Neutral strangelets can then propagate rectilinearly to the Earth and produce a hotspot in the sky of the angular size of the MC, $\theta_{\rm MC}\sim 14^\circ\,(R_{\rm MC}/25\,{\rm pc})(d_{\rm MC}/200\,{\rm pc})^{-1}$, with $d_{\rm MC}$ the distance of the MC to the observer. Note that this corresponds roughly to the size of the observed hotspots. 

The MC can radiate a total energy in neutral strangelets of 
$E^{(1)}_{\rm MC}= \eta {E_{\rm ej}} \,{R_{\rm MC}^2}/{[l(d_{\rm s-MC})]^2}$, 
with $l(d_{\rm s-MC})=d_{\rm s-MC}^2c/(2D)$, the effective distance travelled by a diffusing particle over the rectilinear distance $d_{\rm s-MC}$, separating the center of the source to the center of the MC. The factor $\eta$ is a free parameter that accounts for  strangelet production rate at the source, and the low strangelet neutralization rate in the MC. This expression is only valid for $d_{\rm s-MC} > R_{\rm MC}$. If the source is located at the center of a MC, all the produced particles diffuse in the cloud, and $E^{(2)}_{\rm MC}=\eta E_{\rm ej}$. The total energy of neutral strangelets radiated by the MC can then be expressed over the whole range of $d_{\rm s-MC}$ as $E_{\rm MC}=[1/E^{(1)}_{\rm MC}+1/E^{(2)}_{\rm MC}]^{-1}$.

The excess signal in a solid angle $<\Omega$ around one source can be defined as the following signal-to-noise ratio:
$\sigma_{<\Omega} = N_{\rm s,<\Omega}/(N_{\rm iso,<\Omega})^{1/2}$,
where $N_{\rm s,<\Omega}=
{L_{\rm MC}}A(\alpha,\delta){4\pi d_{\rm s-MC}^2\Omega E}^{-1}
$ 
indicates the number of events expected in a solid angle $<\Omega$ from a source and $N_{\rm iso,<\Omega}=  E{J_{\rm iso,sr}}A(\alpha,\delta)$ the corresponding number of events expected for an isotropic background. For a MC located at coordinates $(\alpha,\delta)$, at distance $d_{\rm MC}$, and separated by $d_{\rm s-MC}$ from the source, the signal at energy $E$ can then be estimated as: 
\begin{equation}\label{eq:sigma}
 \sigma(E)=\frac{\eta}{E^{3/2}}\left[1+\frac{ d_{\rm s-MC}^4c^2}{4D^2R_{\rm MC}^2}\right]^{-1}\frac{E_{\rm ej}}{\Delta t}\frac{A(\alpha,
\delta)^{1/2}}{4\pi d_{\rm MC}^2\Omega J_{\rm iso,sr}^{1/2}}\ ,
\end{equation}
where $A(\alpha,\delta)$ [in m$^{2}$~s~sr] is the exposure of an experiment in the direction $(\alpha,\delta)$, $J_{\rm iso,sr}(E)$ is the observed cosmic ray flux at energy $E$, per steradian, and $\Delta t$ is the diffusion time for particles to travel over a distance $\min(R_{\rm MC},2R_{\rm MC}+d_{\rm MC})$. For a source located inside the MC, the luminosity in neutral strangelets radiated by the MC at $E=20\,$TeV is of order $L_{\rm MC}=E_{\rm MC}/\Delta t\sim 3.5\times10^{40}\,\eta Z^{-1/3}(R_{\rm MC}/25\,{\rm pc})^{-2}\,{\rm erg/s}$. 

\begin{figure}[tp]
\begin{center} 
\includegraphics[width=\columnwidth]{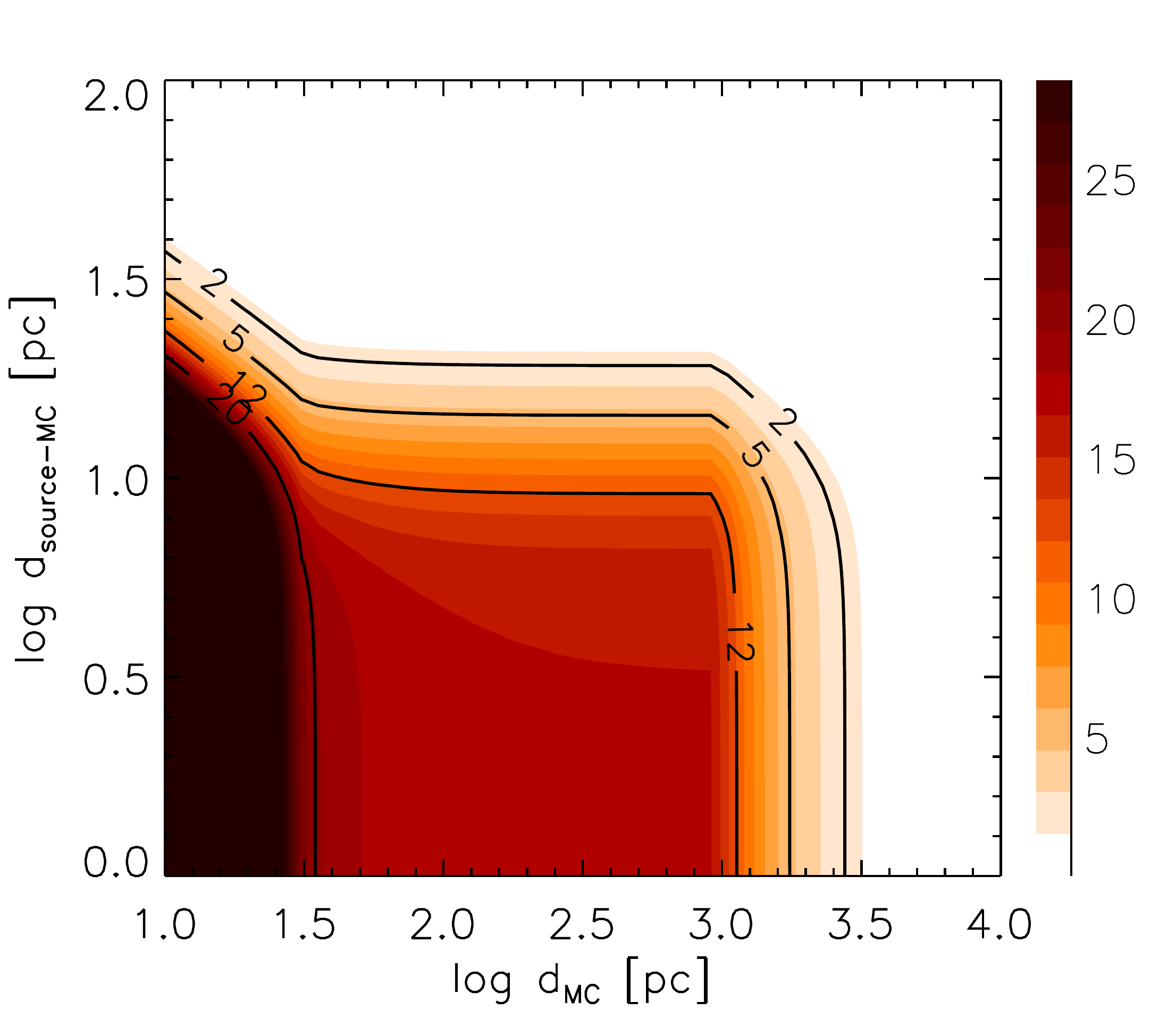}
\caption{Particle excess significance $\sigma$ (Eq.~\ref{eq:sigma}), as would be observed by Milagro with 7 years of data, at $E=20\,$TeV, as a function of  the distance of the MC to the Earth, $d_{\rm MC}$, and the distance between the source and the MC, $d_{\rm s-MC}$, for strangelets with $Z=1$, $A=10^3$ and a MC of radius $R_{\rm MC}=25\,$pc, source of luminosity $L_{\rm MC}=\eta10^{40}\,$erg/s, and an efficiency factor $\eta = 5\times 10^{-8}$. Color bar: value of $\sigma$, black lines: specific numerical values of $\sigma$ as indicated. }
\label{fig:sigma}
\end{center}
\end{figure}

Figure~\ref{fig:sigma} presents contours of the value of $\sigma$ (Eq.~\ref{eq:sigma}) for strangelets with $Z=1$, $A=10^3$ at $E=20\,$TeV, as a function of  $d_{\rm MC}$ and $d_{\rm s-MC}$, for a MC of radius $R_{\rm MC}=25\,$pc. For each set of distances $(d_{\rm s-MC},d_{\rm MC})$, the signal $\sigma$ is calculated for a solid angle $\Omega$ corresponding to an angle in the sky of $\min(3^\circ,\theta_{\rm MC})$. This takes into account the minimum smoothing angle of the anisotropy analysis conducted by Milagro and IceCube ($2.1^\circ/\cos(\delta)$ for Milagro \cite{neutrino_exp} and $\sim 3^\circ$ for IceCube \cite{Abbasi11}). Cosmic ray measurements indicate $J_{\rm iso,sr}(20\,{\rm TeV})\sim5\times 10^{-17}$\,eV$^{-1}$~{s}$^{-1}$~m$^{-2}$ sr$^{-1}$, and we chose an exposure of $A(\alpha,\delta)= 10^{13}\,$m$^2\,$s\,sr, roughly corresponding to the Milagro exposure  at $20\,$TeV, over 7 years of operation.

In our calculation, we set the efficiency factor to $\eta = 5\times 10^{-8}$. Efficiencies in the range of $10^{-8}<\eta<10^{-7}$ lead to reasonable values in terms of $\sigma$ (as $\sigma\propto \eta$), whatever the relative location of the source and the MC, and the distance to the MC. From Eq.~\ref{eq:sigma}, one can infer the strong dependency of $\sigma$ on the distance between the source and the MC: $\sigma\propto d_{\rm s-MC}^{-4}$, when the source is at the border of the MC.  On the other hand, Fig.~\ref{fig:sigma} shows that the value of $\sigma$ is relatively constant as long as the source is at a relatively central position inside the MC. This range of $\eta$ thus implies that only MC within $1-2\,$kpc, and only sources {\it located inside the MC} can produce a significant hotspot (note also that local MCs are found beyond $d_{\rm MC}\gtrsim 70\,$pc). 

Such low values of this effective parameter $\eta$ leave room for combined uncertainties in possibly low strangelet injection at the source, strangelet acceleration, and neutralization efficiencies in the cloud. All these are largely unknown but it is expected that the fraction of strangelet ejected mass at the source should be at most $\sim 1-10\%$ of the mass difference in the transitioning NS and QS configurations \cite{perez-daigne-silk}. The neutralization rates in the cloud should be a small fraction of $r_{\rm spall}$ and $r_{\rm ion}\propto \eta_e$ for spallation and electron capture respectively.

Magnetic fields in MCs are known to scale approximately  with gas density as $n^{1/2}$ relative to the mean Galactic magnetic field. The diffusion coefficient scales in $r_{\rm L}^{1/3}l_{\rm c}^{-2/3}\propto B^{-1/3}l_{\rm c}^{-2/3}$ in the Kolmogorov diffusion regime. Taking into account these stronger fields would thus only result in an order of magnitude difference in $\sigma$, and the variations could be absorbed by the uncertainty in the efficiency factor $\eta$.

Whether the NS-QS transition actually gives birth to a pulsar is unknown. Hence strangelet sources will not necessarily be found at the position of an active source, and one reasonable assumption would be that they are distributed as old NS. In general these are everywhere, including in MC. Indeed, the number density of old NS in the Galaxy is of order $\sim 10^{-4}\,{\rm pc}^{-3}$ \cite{Lyne06}, and if 10\% of them do not get kicks at birth (i.e., remain in the MC), their mean separation is 20 pc. 
Molecular clouds and gamma ray pulsars have a similar distribution on the sky, while radio
pulsars cover a much wider age range  and have a broader distribution. The fraction of old neutron stars that are required to undergo NS-QS transition in order to account for the handful of observed hotspots is $\sim 10^{-4}$. The diffusion time of strangelets in MCs, i.e., the time over which each source will be observable, is of order $\Delta t\sim Z^{1/3}\,375\,$yrs. This implies a NS-QS transition rate of order $3\times 10^{-7}\,{\rm yr}^{-1}$.

\section{Comparison with data and signatures}

Most observed hotspots could be produced by MCs in the Gould Belt (a star forming region concentrating many MCs, that forms a ring at a distance from the Sun of $\sim 0.7-2\,$kpc), at the location where NS-QS transitions may have occurred. Interestingly, the Milagro hotspot labelled ``Region A" \cite{neutrino_exp} lies in the direction of the Taurus Molecular Cloud, the nearest star formation region located at 140\,pc, and that covers $\sim 100\,{\rm deg}^2$ in the sky \cite{clouds}. ``Region 1" of IceCube \cite{Abbasi11} is also in the direction of a remarkable MC: the Vela Molecular Ridge, located at $0.7-2\,$kpc distance, of size $\sim 15^\circ$ in sky \cite{clouds}. 

It is difficult to predict whether strangelets could produce air-showers conspicuously different from those from ordinary cosmic rays, mainly because their cross-section is not known. Preliminary hadronic simulations with EPOS and CONEX show that the strangelet heavy mass (expected to produce shallow showers with low fluctuations) and its high binding energy (with the oposite effect) could also compensate each other to produce ordinary cosmic ray showers \cite{Schuster12}.
It leaves room for the possibility that the cosmic rays observed by \cite{neutrino_exp,Abbasi11} actually be strangelets, that were not identified (as long as $A\lesssim 10^2-10^4$, see earlier discussion). IceCube also reported composition measurements from 1 to 30 PeV, that are compatible with ordinary cosmic rays \cite{Abbasi12}. 
The muon rate for strangelets is expected to increase in $\propto A^{0.1}$, which should not have a noticeable impact for our values of $A$. This again is consistent with the IceCube reports. 

Finally, our model predicts  point-like cosmic ray sources farther than a few kpc.
In Fig.~\ref{fig:sigma}, the cut-off of the signal at distances $d_{\rm MC}\gtrsim 1-2\,$kpc stems from the limited resolution angle and sensitivity of the instruments, which set the smoothing angle for the anisotropy search. Indeed, as the angular size of the MC in the sky diminishes, the signal is diluted inside one angular bin, and cannot be distinguished from the noise. With better resolution and sensitivity (with the High Altitude Water Cherenkov Observatory, HAWC, for example), the excess signal could remain high at larger distances, and point-like sources could be spotted. 

This scenario could also lead to multi-messenger signatures in secondary neutrinos or gamma-rays. These particles could be produced directly at the source when strangelets are generated, or when particles undergo spallation in the clouds. In the former case, the secondary signal should be point-like. In the latter case, the whole MC could be illuminated, and
the expected gamma-ray flux should be comparable to the product of a multiplicity factor multiplied by the hot spot flux. The multiplicity factor is of the order of the spallation fraction in the MC or $r_{\rm spall}\simlt 10.$ The predicted flux over solid angle $\Omega$ sr is
$N_\gamma(>1{\rm TeV}) \sim 5.10^{-12} (\sigma/10)(r_{\rm spall}/10)(\Omega/10^{-5})\,{\rm s^{-1}cm^{-2}}.$
This should be compared with the expected flux of cosmic ray induced gammas.\\

We discussed the possibility that strangelets accelerated in nearby NS-QS transitions, and then becoming neutral by spallation or electron capture in molecular clouds, could explain the small-scale anisotropies observed by several experiments at TeV-PeV energies. With reasonable astrophysical assumptions regarding NS-QS transition rates, particle injection and neutralization rates, we can reproduce successfully the observed hotspot characteristics and their distribution in the sky. \\

We thank T. Delahaye, T. Pierog and A. Olinto for fruitful discussions. We thank the COMPSTAR and MULTIDARK projects, Spanish MICINN projects FIS-2009-07238 and FIS2012-30926. M.A.P.G would like to thank the kind hospitality of IAP where part of this work was developed. K.K. acknowledges support from PNHE.



\begin{thebibliography}{9}
%

\bibitem{neutrino_exp} G. Guillian et al., PRD 75, 062003 (2007); M. Amenomori et al., Science 314, 439 (2006); 
A. A. Abdo et al., PRL 101, 221101 (2008); A. A. Abdo, et al., ApJ 698, 2121 (2009); S. Vernetto, Z. Guglielmotto, J. L. Zhang, and for the ARGO-YBJ Collaboration, ArXiv: 0907.4615 (2009); R. U. Abbasi et al., PRL 104, 161101 (2010)
\bibitem{Abbasi11} R. Abbasi, et al., ApJ 740, 16 (2011)
\bibitem{large_scale_theories} A. D. Erlykin and A. W. Wolfendale, Astroparticle 25, 183 (2006);  P. Blasi and E. Amato, JCAP 1, 11 (2012)
\bibitem{previous_works} L. O. Drury and F. A. Aharonian, Astroparticle Phys. 29, 420 (2008); 
M. Salvati and B. Sacco, A\&A 485, 527 (2008); G. Giacinti and G. Sigl (2011), ArXiv:1111.2536
\bibitem{Han08} J. L. Han, Nuclear Physics B Proc. Suppl. 175, 62 (2008)
\bibitem{witten84} Witten, E., Phys. Rev. D, 30, 272 (1984)
\bibitem{Perez-Garcia} M. A. Perez-Garcia, J. Silk, and J. R. Stone, PRL 105, 141101 (2010); M. A. Perez-Garcia and J. Silk, ArXiv:1111.2275
\bibitem{merger} R. Oechslin, K. Uryu, G. Poghosyan, F. K. Thielemann, Mon. Not. Roy. Astron. Soc., 349, 1469, (2004)
\bibitem{Wilk96} G. Wilk and Z. Wlodarczyk, Journal of Physics G Nuclear Physics 22, L105 (1996)
\bibitem{Madsen}  J. Madsen, PRL 85, 4687 (2000); J. Madsen, PRL 87, 172003 (2001); J. Madsen, PRD 71, 014026 (2005); J. Madsen, arXiv:0612740;  J. Madsen, arXiv:0512512
\bibitem{strangelet_accelerators} J. Madsen, PRD 71, 014026 (2005); K. S. Cheng and V. V. Usov, PRD 74, 127303 (2006)
\bibitem{strange_stars} C. Alcock, E. Farhi, and A. Olinto, ApJ 310, 261 (1986); C. Alcock and A. Olinto, ARNPS 38, 161 (1988)
\bibitem{Perez-Garcia12} M. A. Perez-Garcia, F. Daigne, and J. Silk, arXiv:1211.7018 
\bibitem{Padovani09} M. Padovani, D. Galli, and A. E. Glassgold, A\&A 501, 619 (2009) 
\bibitem{cosmic_rays_pulsars} W. Bednarek and R. J. Protheroe, PRL 79, 2616 (1997); P. Blasi, R. I. Epstein, and A. V. Olinto, ApJ Letters 533, L123 (2000); J. Arons, ApJ 589, 871 (2003); K. Kotera, PRD 84  (2011) 023002, K. Fang, K. Kotera, and A. V. Olinto, ApJ 750, 118 (2012)
\bibitem{perez-daigne-silk}M. A. Perez-Garcia, F. Daigne, and J. Silk, submitted
\bibitem{Blasi12_1} P. Blasi and E. Amato, JCAP 1, 10 (2012)  
\bibitem{Lyne06} A. G. Lyne and F. Graham-Smith, {\it Pulsar Astronomy}, Cambridge University Press (2006)
\bibitem{clouds} G. Narayanan, M. H. Heyer, C. Brunt, P. F. Goldsmith, R. Snell, and D. Li, ApJ S. 177, 341 (2008); D. C. Murphy and J. May, A\&A 247, 202 (1991)
\bibitem{Schuster12} D. Schuster and L. Wiencke, in APS April Meeting Abstracts (2012), p. 7007
\bibitem{Abbasi12} R. Abbasi, et al. 	(2012), ArXiv:1207.3455
\end{thebibliography}
\end{document}